\documentclass[12pt,preprint]{aastex}
\usepackage{graphicx}
\usepackage{enumerate}
\usepackage{amssymb}
\usepackage{amsmath}

\shorttitle{Dark Energy and Black Holes}
\shortauthors{Aykutalp \& Spaans}

\begin{document}

\title{Testing the Proposed Connection between Dark Energy and Black Holes}

\author{A.\ Aykutalp\altaffilmark{1} and M.\ Spaans\altaffilmark{1}}
\affil{Kapteyn Astronomical Institute, University of Groningen, PO Box 800, 9700 AV Groningen, The Netherlands}

\email{aykutalp@astro.rug.nl}
\email{spaans@astro.rug.nl}

\begin{abstract}
In 1997, an extension of general relativity was
proposed that predicts the dark energy density $\Lambda$ to vary linearly with
the total number of macroscopic black holes in the entire three-volume of
the universe. We explore this prediction and find
that $\Lambda$ must be roughly constant after the bulk of the stellar mass
black holes are in place, so for a redshift $z$ smaller than unity.
Conversely, the highest black hole formation rate corresponds to the peak in
the cosmic star formation history, earlier than $z=1$. This yields a fast
declining $\Lambda$, by a factor of about 5, from redshift 1 to 3.
At even earlier times, before many stars were formed, the value of $\Lambda$
should be much smaller than its current value.
These predicted effects are all consistent with current data, and near
future observations can definitively confirm or disproof the link between
the dark energy density and the total number of black holes in the entire
universe.
\end{abstract}


\section{Introduction: Dark Energy and Theories of Gravity}
The expansion of the universe appears to be accelerating, as indicated by
the magnitude-redshift relation of type Ia supernovae \citep{2010ApJ...716..712A, 1999ApJ...517..565P,  1998Natur.391...51P, 1998ApJ...509...74G, 1998ApJ...507...46S, 1998AJ....116.1009R}.
This phenomenon is usually referred to as dark energy and constitutes one
of the big mysteries in cosmology. Dark energy seems to be fundamentally
connected to the dynamics of space-time itself. The simplest expression of
dark energy is in the form of a positive cosmological constant term on the
left hand (geometric) side of the Einstein equation. The origin of such a
constant is not specified by general relativity (GR), though, leaving one
with an incomplete description of space-time dynamics.

The success of Einstein gravity is overwhelming. Hence, any alternative
theories should take great care in preserving the beautiful features of GR,
while pursuing the mystery of dark energy.
Often, creative and compelling ideas notwithstanding, ad hoc elements or
unconstrained parameters and fields are part of modified gravity theories,
e.g., vector-tensor theory,
quintessence, M-theory, Einstein-aether theory, MOND
\citep{2001PhRvD..64b4028J, 2010PhRvD..82l9901J, 2007PhRvD..75d4017Z, 2009PhRvD..79h4008H, 2011PhRvD..84h4024S}.

An extension of GR has been proposed in \cite{1997NuPhB.492..526S}, S97
from hereon. This theory leads to a unique and testable prediction for dark
energy, while fully preserving GR, without the introduction of any additional
degrees of freedom. The S97 paper is somewhat technical (it uses the
mathematics of algebraic topology) and we therefore provide an extensive
summary of its salient physical properties below.

\section{Physical Aspects of the S97 Theory}
The Einstein equation has many symmetries, but it is not invariant under
local conformal (scale changing) transformations.
As a result, one finds larger metric fluctuations when one goes to smaller
spatial scales. In fact, at the Planck scale of $l_{\rm P}\sim 10^{-33}$ cm,
these fluctuations in space-time become huge (even singular) and space-time
should take the form of a quantum foam with Planck mass
($m_{\rm P}\sim 10^{-5}$ g)
black holes (BHs), so-called mini BHs, popping out off and into the
vacuum \citep{1957AnPhy...2..604W}.
Space-time thus enjoys the presence of mini BHs at the Planck scale, i.e., is
multiply connected. A fundamental problem with the quantum foam is that it
fluctuates violently on a Planck time of $t_{\rm P}\sim 10^{-43.5}$ sec
and is therefore difficult to conceive as a stable structure.
It is important to realize here that GR is a local geometric theory.
Hence, it allows for freedom in the (global) topology of space-time.

S97 seeks stability for the quantum foam through GR, which is the natural low
energy limit of any quantum gravity theory. Specifically, the Schwarzschild
solution for a macroscopic BH can be used as the low energy limit of the
quantum foam in a topological sense. That is, as a low energy limit that
allows for changes in the geometry of space-time but one that also assures all
local observers to count the same number of event horizons. I.e., the
connectivity of space-time, so the topology of the event horizon, does not
change if one makes a BH larger. Furthermore, such a countable quantity like
the number of BHs is invariant under continuous deformations
(homeomorphisms) of space-time, which is the essence of topology, unlike
geometry, which deals with differentiable transformations (shape preserving
diffeomorphisms).

This kind of a low energy limit uses the fact that all BHs can be described
geometrically by just their mass, charge and spin, while they all possess a
closed surface as well (the event horizon), irrespective of their local
geometry. Furthermore, BHs have a black body temperature \citep{1975CMaPh..43..199H} and
small BHs are much warmer than large BHs (the effective temperature goes as
1/mass). However, mini BHs in the quantum foam are transient in that they are
continuously created and destroyed. Hence, if GR is to be a proper low energy
limit of the quantum foam (with a well defined number of event horizons for all
observers), the existence of mini BHs must somehow be subject to the presence
of macroscopic BHs. Afterall, only the latter can exist (much) longer than a
local Hubble time under Hawking evaporation.

In S97, the quantum foam stability problem that plaques GR at the
Planck scale is solved by allowing only macroscopic (long lived)
BHs to {\it induce} mini BHs in every local Planckian volume
$\sim l^3_{\rm P}$. One needs to consider distinct Planckian
volumes here since quantum physics forces space-time to be
quantized at the Planck scale if one considers that observers
cannot measure with a better accuracy than $\sim l_{\rm P}$.
The S97 argument for the induction of mini BHs is simple:
Mach's principle, as interpreted by Einstein, dictates that
the global distribution of matter determines local geometry.
Wheeler's local quantum foam, with its non-trivial topology in the
form of mini BHs, is also a direct expression of GR. Therefore,
following Einstein's ideas for geometry, matter that is globally
in the form of macroscopic BHs must determine the Planck scale
topology of GR.

Topology, with its invariance under continuous deformations of
space-time, cares only about a countable quantity like the number
of BHs in the entire three-volume of the universe (like the number
of handles on a mug).
Hence, the number of macroscopic BHs in the entire universe,
a global quantity, determines the local occurrence of mini BHs.
To be more quantitative about this, one must compute how many mini
BHs Wheeler's quantum foam enjoys in every Planckian volume, for a
specific number of macroscopic BHs. I.e., one needs to assess with what,
possibly time dependent, probability $\delta$ these induced mini BHs are
created in a Planckian volume. The key aspect of $\delta$ is therefore how it
depends on the total number of BHs in the entire three-volume of the universe.

Following S97, one first notes
that any topologically non-trivial space-time manifold can be constructed
by adding together three-dimensional building blocks called primes. These
primes (e.g., handles, three-tori, three-spheres) lend their name from the
property that they themselves cannot be written as the connected sum of
smaller three-dimensional units. Furthermore, interactions among those
primes can never change their intrinsic properties (like the number of
closed loops they contain). Finally, following the superposition principle
of quantum mechanics, any property of the quantum foam must be carried by
the linear sum of these primes. The topological extension of Mach's
principle above then states that the total number of macroscopic BHs in the
universe determines the properties of the quantum foam. Macroscopic BHs are
primes (half BHs do not exist) and have the same intrinsic topological
properties as mini BHs in the quantum foam. These macroscopic BHs are a
linear measure of the global space-time topology, as the connected sum of all
those BHs, just like the mini BHs in the quantum foam are such a linear
measure on the Planck scale.
Hence, $\delta$ is {\it linearly} proportional to the number of macroscopic
BHs in the entire three-volume of the universe.

This, currently tiny, probability $\delta$ for a single Planckian volume to
enjoy the temporary presence of mini BHs is computed below, but its crucial
property is that it is linear in the number of macroscopic BHs (so counting
event horizons) and independent of geometry (so volume).
The latter property forces all local Planckian volumes in the quantum foam to
respond as one to the presence of macroscopic BHs, irrespective of
whether the universe is expanding (and thus adding Planckian volumes).
I.e., dark energy is {\it purely topological} in nature and every observer
witnesses the same local topological changes in the Planckian quantum foam
(so space-time has uniform statistics), subject only to how the low energy GR
limit drives the formation of macroscopic BHs globally.

One may think that this almost instantaneous response (within a time slice) of
the dark energy everywhere to BH formation leads to causality violations, but
this is not the case:
In S97, the equations of motion for the topology of the universe are derived.
These equations contain no spatial information and are first order in the
discretized time variable $t=nt_{\rm P}$, for integers $n\ge 1$ that
distinguish different time slices of thickness $t_{\rm P}$.
I.e., for any time $t_1$ labelling the complete three-space of the universe,
the solution one Planck time later only depends on the global topology at time
$t_1$. Also, under a continuous
transformation one can always deform three-space in such a manner that any two
three-space points are brought within each other's vicinity. Subsequently,
information on the creation of any BH, which merely amounts to raising the
count of event horizons by one, can thus be brought into the vicinity of any
local observer as well. Furthermore, this freedom of continuous three-space
deformation assures that no such information
needs to be exchanged faster than the speed of light to inform every local
Planckian volume, no matter how distant from the BH, of the topology changing
event. After all, the distance between the two aforementioned three-space
points can be made as small as a Planck length by the deformation.
One should realize here that BHs and observers are not moved physically
during such continuous deformations. Rather, it is space-time itself that
deforms and facilitates the transfer of BH number information through
topological identifications between space-time points.

The latter can best be imagined as loops that provide short-cuts through
four-space between any two points in three-space.
These loops can be constructed by taking a line segment (a path) and
identifying two points along it to create a loop as an alternative route.
Doing this for three dimensions, one constructs a three-torus where each
surface of the cube connects to the opposite one through the fourth dimension
(time).
Along these short-cuts information travels at maximally the speed of light,
obeying local Lorentz invariance, but across arbitrary three-space distances.
Creating these short-cuts through four-space between all pairs of
Planckian volumes allows topological information to be exchanged causally
within {\it the same} given time slice of thickness $t_{\rm P}$. The resulting
structure is a lattice of three-tori (see S97 for further details).
Hence, wherever a BH is formed in some time slice its effect on the dark energy
density in that time slice is transferred purely topologically through these
identifications, without the need for causality violations and expressed
solely by an update on the total number of BHs. This testable
topological aspect of dark energy augments the local geometric
approach of GR, and fully preserves the latter, while it assigns countable
numbers (BH horizons) to entire time slices.
Time is relative of course and one may freely choose any foliation for the
time slices since the number of BHs is a covariant quantity.

Because topology allows one to identify all space-time points as above, it
is the total number of macroscopic BHs in the entire universe, so in all
of three-space for every time slice, that is pertinent to the dark energy
density. Therefore, the
spatial distribution of macroscopic BHs, or the actual volume of our evolving
universe, is irrelevant.
In all, S97 finds the dark energy density $\Lambda$ to be linearly
proportional to the total number of macroscopic BHs in the entire three-volume
of the universe, $N_{\rm BH}(z)$, at a given redshift $z$.
In this, redshift is a distinctive measure of time in the usual $3+1$
decomposition for the evolution of the three-geometry and three-topology
of space-time. I.e., redshift labels different global time slices in the
geometrical and topological dynamics of three-space. Algebraically,
\begin{equation}
\Lambda (z)= \delta (z)\frac{m_{\rm P}}{l^3_{\rm P}},\label{eq:no1}
\end{equation}
where
\begin{equation}
\delta (z)=N_{\rm BH}(z) \frac{l^3_{\rm P}}{l^3_{\rm I}},
\end{equation}
and $l_{\rm I}$ the size of the universe (after inflation, marking the
transition to the low energy GR limit) when the first BH forms that exists for
longer than a local Hubble time. I.e., $l_{\rm I}$ is {\it frozen in} when the
quantum foam stabilizes and each induced mini BH contributes about
$m_{\rm P}\sim 10^{-5}$ g of mass per Planckian volume $l^3_{\rm P}$.
Equation \ref{eq:no1} shows that, for $N_{\rm BH}(z)$ independent of redshift,
any local observer concludes that the dark energy density $\Lambda$ in g
cm$^{-3}$ remains constant as the universe expands.
Conversely, the dark energy density scales linearly with the total number of
macroscopic BHs that are present in the entire three-volume of the universe at
a given redshift.
Induced mini BHs can drive the expansion of the universe, so constitute a
negative pressure, because their creation as three-dimensional spatial objects
requires an increase in four-space volume to actually embed them. I.e., every
formed event horizon is a closed surface to an external observer.

The absolute value of $\Lambda$ is only roughly estimated in S97 due to
uncertainty in $N_{\rm BH}$ today and $l_{\rm I}$.
However, the exact value of $\Lambda$ is not important here because this paper
concentrates only on the evolution of the dark energy density with the
total number of BHs in the entire universe. Of course, the observable
universe may be smaller than the entire universe, which is relevant for
measuring the total number of BHs through observations. If we assume that the
observable universe is
the entire universe, then one can estimate $l_{\rm I}$ as follows.
Today's dark energy density is $\sim 10^{-29}$ g cm$^{-3}$ and the universe
presently contains $10^{19}$ macroscopic BHs. The latter
number follows if one assumes that 1 in $10^3$ stars yields a BH, for a
Salpeter initial mass function (IMF), and that there are $10^{11}$ galaxies
with each $10^{11}$ stars. Hence, with these numbers $l_{\rm I}\sim 2\times 10^{14}$ cm.

The S97 theory provides solutions for the evolution of the very early
universe's topology.
The number of mini BHs increases exponentially during the quantum gravity 
phase of the universe (so before the first macroscopic BH is formed and the 
quantum foam is stabilized). Therefore, mini BHs can drive an inflationary
period  at these early unstable times, with a dark energy density that
corresponds to about one mini BH per Planckian volume ($\delta\sim 1$).

In all, one has the local space-time quantum foam
that \citep{1957AnPhy...2..604W} envisioned, but globally constrained by
the total number of macroscopic BHs in the entire universe.
The value of $N_{\rm BH}(z)$ thus controls
the temporal variations in the dark energy density, in the spirit of Mach's
principle, where global properties of the universe determine local ones.

\section{Testing the Connection between Dark Energy and Macroscopic Black Holes}
The universe appears to be approximately flat based on WMAP7 cosmological
observations \citep{2011ApJS..192...18K} and $\Lambda$ is about
$0.734\pm 0.029$ in
units of the critical density $\rho_c=1.88\times 10^{-29}$ g cm$^{-3}$. As
discussed above, the dark energy density of our universe is linearly
proportional to the total number of macroscopic BHs in three-space at any
redshift. Very practically, this means that if the number of BHs in the
entire universe at redshift 1 is twice higher than at redshift 2, then
$\Lambda$ in g cm$^{-3}$ doubles from $z=2$ to $z=1$. 
This paper focuses on the putative evolution in $\Lambda$. Therefore,
it is only necessary to compute how the total number of BHs in the universe
changes with redshift, irrespective of what the current value
of $\Lambda$ is.

The bulk of the stars, and thus stellar BHs in the universe appears to be
present as
early as $z=1$, when a $(1+z)^4$ decline ensues in the cosmic star formation
rate \citep{2006ApJ...651..142H}.
This is earlier than current type Ia supernova (SN) detections probe, hence
one expects to find an effectively constant $\Lambda$ for $z<1$.
It is relevant here that type Ia SNe occur with a delay of close to 3 Gyr
relative to regular star formation.
Conversely, given the rapidity with which massive stars/BHs are produced
around $z=1-3$, this epoch should exhibit a strong decrease in the number of
BHs, and thus $\Lambda$, from $z=1$ to $z=3$.

These qualitative expectations can be quantified as follows.
\cite{2006ApJ...651..142H}, their figure 7, derive the comoving type II SN rate
density from the star formation history of the universe.
These authors adopt a type II SN mass range of 8-50 M$_\odot$ and the ``BG''
and ``SalA'' (Salpeter) IMFs of \cite{2003ApJ...593..258B}.
We use their $z=0-6$ data for a Salpeter IMF to estimate the evolution of the
dark energy density
from today's value back to higher redshift, in terms of the comoving type II
SN rate density $dn_{\rm SN}/dt$ (in yr$^{-1}$ Mpc$^{-3}$).
In this, we assume that stellar mass BHs are produced by type II SNe with some
constant efficiency $\epsilon$.
The comoving type II SN rate density is used here because such an
expansion-corrected coordinate system allows one to approximate the
relative change in the total number of BHs that are formed in the
universe at a given redshift, as follows. The used data, and thus the
derived evolution of $\Lambda (z)$, pertain to the observable universe only.
Hence, we assume that the observable universe gives a good statistical
representation of
the entire universe as far as the change in the total number of
macroscopic BHs in evolving three-space is concerned.
We thus seek $\Lambda (z)$
normalized to its present value $\Lambda (0)$, which corresponds to the
current number of BHs in the observable universe.

\begin{figure}[!htb]
\begin{center}
\includegraphics[angle=0,width=12cm]{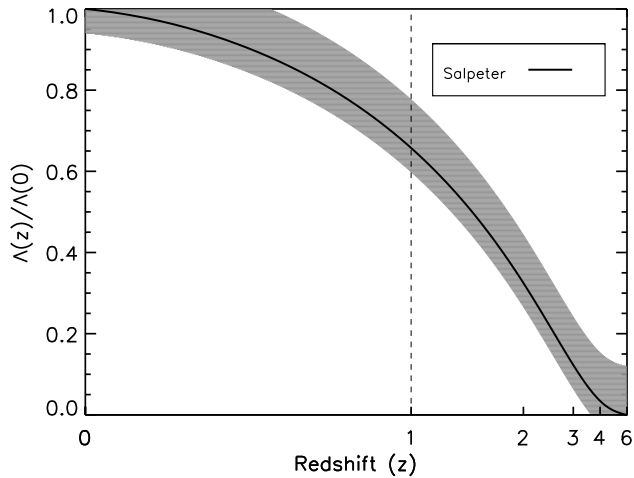}
\caption{Redshift evolution of $\Lambda$ derived from the type II SN rate data of \cite{2006ApJ...651..142H}.
The shading represents the observational uncertainty in deriving the type II SN rate.}
\label{fig:DE}
\end{center}
\end{figure}

The ratio $\Lambda (z)/\Lambda (0)$ can now be expressed analytically, in
terms of $dn_{\rm SN}/dt$ for observers that measure the rate of type II SNe
in their comoving volume, as
\begin{equation}\label{eq:DE}
\frac{\Lambda (z)}{\Lambda (0)}=\frac{\int_6^z dt(z)\ \epsilon\ dn_{\rm SN}/dt}{\int_6^0 dt(z)\ \epsilon\ dn_{\rm SN}/dt},
\end{equation}
and is independent of the constant BH formation efficiency $\epsilon$.
We perform the integrals in equation \ref{eq:DE} for a flat cosmology and the
same cosmic time interval as used by \cite{2006ApJ...651..142H}, 
$t(z) = \frac{1}{H_0} \int_0^z \frac{dz}{(1+z)[\Omega_m (1+z)^3 + \Omega_{\Lambda}]^{\frac{1}{2}}}$.
We use redshift as a coordinate, but there is no pertinent dependence of our
results on coordinate system because of the normalization in \ref{eq:DE}. Any
observer, whatever its coordinate system, will find the dimensionless ratio of
two numbers to be covariant. One
may worry about BHs that are formed beyond our horizon. However, as explained
above, these BHs do contribute to the dark energy density in our comoving
volume, but without violating causality.
At the same time, the right hand side of \ref{eq:DE} is empirical.
So, even though we, as observers, have a special perspective on (the history
of) the universe in our current time slice, we can still derive the change in
the total number of BHs, as three-space evolves, if we live in a representative
comoving volume. I.e., one that enjoys the same time dependence of BH
formation as the whole universe.
Figure \ref{fig:DE} shows the evolution of $\Lambda (z)/\Lambda (0)$ and
constitutes a quantitative prediction of the (parameter free) S97 model.

It is found that $\Lambda (z)/\Lambda (0)$
changes by less than $\sim 30$\% more recently than a
redshift of unity. Furthermore, a rapid decline in the dark energy density by
a factor of 5 occurs from $z=1-3$. The shading in figure \ref{fig:DE}
represents the observational uncertainty in deriving the type II SN rate from
the cosmic star formation history \citep{2006ApJ...651..142H}.
For redshifts $z>5$ the shading indicates that a significant amount of star/BH 
formation may occur at very early times (but see the discussion section).
The value of the equation of state parameter, $w=\frac{p}{\rho}$ 
(for pressure $p$ and density $\rho$) is estimated to be
$-1.023\pm 0.09 {\rm (stat)} \pm 0.054 {\rm (sys)}$ \citep{2010ApJ...716..712A, 2006A&A...447...31A} and $-1.061^{+0.069}_{-0.068}$ \citep{2011ApJ...737..102S}, for $z<1$.
In figure \ref{fig:DE}, we see that $\Lambda (z)/\Lambda (0)$ is roughly constant and slightly
decreasing with increasing redshift for $z<1$. This is consistent with
observational data, because a constant $w = -1.1$ corresponds to an
allowed change in $\Lambda (z)/\Lambda (0)$ of about 30$\%$. The latter follows
from the conservation equation
\begin{equation}
{{d\rho /dt}\over{\rho}}=-3H(1+p/\rho )=-3H[1+w(z)],
\end{equation}
with the Hubble parameter squared $H^2=8\pi\rho /3$ \citep{2003PhRvL..90i1301L}.
If one expands $1+w$ as $1+w\approx \delta w$, the relative change
in $\rho$ over a local Hubble time is about
$(H\rho )^{-1}d\rho /dt=-3\delta w$, for constant $\delta w$.
Indeed, figure \ref{fig:DE} shows that $\Lambda$ decreases by about 30\%,
for a constant $w\approx -1.1$, from $z=0$ to $z=1$. Furthermore, the sharp
drop in $\Lambda (z)$ for $z>1$ then shows that no constant value of $w$ can
mimic dark energy behavior for $z\sim 0-2$ if the total number of stellar
mass BHs controls the vacuum energy density.
Also, \cite{2010ApJ...716..712A} point 
out that although $-w$ is close to unity there is room for evolution for $z>1$.
This is
because the SN Ia data only weakly constrain dark energy for such early times.
Similar constraints are provided by \cite{2005PhRvD..71j3515S}, based on SNe,
SDSS and CMB data, as well as by \cite{2011arXiv1109.6125L}, based on SN legacy
survey data.

Furthermore, recent WMAP7 work by \cite{2011ApJS..192...18K}, provides limits
on the properties of time dependent dark energy parameterized by
$w(a)=w_0+w_a(1-a)$ \citep{2001IJMPD..10..213C, 2003PhRvL..90i1301L},
where $a$ is the scale factor ($a=\frac{1}{1+z}$). In their table 4, these
authors estimate $w_0=-0.93\pm 0.12$ and $w_a=-0.38^{+0.66}_{-0.65}$,
for a flat universe. Their 13\% uncertainty in $w_0$ is easily accommodated
in our figure \ref{fig:DE}, while the fiducial range of $w_a$ allows for the
factor of 5 change in $\Lambda$ over $z=1-3$. Recent results by \cite{2011ApJ...737..102S}
provide $w_0=-0.905\pm 0.196$ and $w_a=-0.984^{+1.094}_{-1.097}$,
which are again consistent with our figure 1.
Indeed, because the bulk of the stellar mass BHs appears not to have formed
before $z=1$, we expect to have $w<-1$ for $z>1$. Nevertheless, it is also
apparent that a modest improvement in future measurement accuracy around
$z\sim 1$ could confirm or rule out the S97 model.

\section{Discussion}
One can estimate the number of BHs at $z\sim 6$ from the soft X-ray background
(0.5-2 keV), which is partly produced by accreting black holes at $z>6$.
This is relevant because the Hopkins \& Beacom (2006) data stop around a
redshift of 6.
\cite{2004ApJ...613..646D} estimate an upper limit for
the BH mass density of $4\times 10^4$ M$_{\odot}$ Mpc$^{-3}$. This translates
into $10^{14}$ BHs for the observable universe at $z\sim 6$, if the Pop III IMF
is dominated by stars of
$10-100$ M$_{\odot}$ \citep{2002ApJ...564...23B, 2002Sci...295...93A}.
This is orders of magnitude smaller than the number of BHs in the observable
universe at $z=0$, $N_{\rm BH}\sim 10^{19}$.
Since the bulk of the stellar mass BHs is not assembled at $z>5$, stellar
origin BHs should play no role in the dark energy density at those early times.

A top-heavy IMF may aid the formation of BHs. Studies by
\cite{2010A&A...522A..24H, 2011A&A...536A..41H} indicate that star formation
within $r_c=100 (M/10^7 {\rm M}_\odot )^{1/2}$ pc of a supermassive
($M>10^5$ M$_{\odot}$) BH, accreting at 10\% of Eddington, leads to a
top-heavy IMF with as much as
5-10\% of all stars formed being more than 8 M$_\odot$. This can provide an
additional boost to the BH production rate at $z>2$ when quasars accrete
rapidly, while also being consistent with the SN neutrino constraints for
$z<1$ \citep{2006ApJ...651..142H}. The total amount of gas involved here is
likely less than 10\% of the total galactic reservoir, given that $r_c$
is generally smaller than 1 kpc. Still, if obscured infrared and sub-mm
bright high-redshift galaxies harbor a large (50\%) fraction of all star
formation in the early universe (for $z\sim 2-3$), then the peak in the cosmic
type II SN rate may be higher and wider than expected \citep{2011MNRAS.415.2723C}.
Subsequently, this would cause a factor of about 2 higher
$\Lambda (z)/\Lambda (0)$ than shown by the grey area in figure \ref{fig:DE}
around $z\sim 3$.

Of course, mergers between BHs may suppress the total number of
macroscopic BHs, and thus the magnitude of $\Lambda$. The merger rate of
stellar mass BHs, which form the bulk of the BH population in the universe
(rather than supermassive ones), 
in BH-BH binaries, is about 500 yr$^{-1}$ \citep{2007PhRvD..76f1504O}.
Given that presently roughly 3 BHs are formed per
second in the observable universe, assuming again that about
0.1$\%$ of all formed stars yield a BH (for a Salpeter IMF), this is unlikely
to decrease $\Lambda$ significantly.

The present number of BHs in the universe could be declining if numerous
primordial black holes were created during the quantum gravity phase of the
universe and are evaporating around this time \citep{1976ApJ...206....1P}. Such
primordial
BHs can survive, and dominate by number today, only if they are more massive
than $M_0\sim 2\times 10^{15}$ g. For $M>M_0$, primordial BHs withstand Hawking
evaporation for more than a local Hubble time since the latter scales
$\propto M^3$.
However, it is likely that the mass function of primordial BHs scales more
steeply than $1/M$. Consequently, the total number of primordial BHs declines
with cosmic time (S97).
This would lead to a strongly diminishing $\Lambda$ with decreasing redshift,
inconsistent with current observational constraints.

High precision measurements of $w$, $w_0$ and $w_a$, as cited in the great
efforts above, will continue to decrease their error bars to the few percent
level for $z<1$ type Ia SNe. They can thus probe the modest but real decrease
in $\Lambda$ shown in figure 1 for $z=0-1$. Furthermore, a combination of
type Ia SNe, galaxy clustering and weak lensing may be quite powerful to
constrain systematics in this \citep{2012arXiv1201.2110}.
Of course, the bulk of the evolution in $\Lambda$ lies at $z>1$.
In this light, it has been shown recently that active galactic nuclei (AGN)
can be used for high redshift (upto $z\sim 4$) measurements of dark energy,
using reverberation mapping \citep{2011ApJ...740L..49W}. This novel, and very
timely, technique is particularly well suited to sample the
$\Lambda (z)/\Lambda (0)$ curve for $z=1-3$, because AGN can be seen to
much higher redshift than type Ia SNe. This method could thus also confirm
or disproof the linear scaling between the dark energy density and the total
number of macroscopic BHs. Of course, because the mean density of the universe
increases as $(1+z)^3$, the relative influence of the dark energy density is
less for $z>1$ and high accuracy is required.

\acknowledgments
The authors thank Andrew Hopkins for sending the type II SNe data derived
from the star formation history of the universe, and Marianne Vestergaard
for discussions on AGN reverberation mapping.

\end{document}